\documentclass[aps,prl,twocolumn,groupedaddress,showpacs]{revtex4-1}

\usepackage{graphicx}
\usepackage{dcolumn}
\usepackage{bm}
\usepackage{epsfig}

\def\be {\begin{equation}}
\def\ee {\end{equation}}
\def\F{{\cal F}
\def\GV{G_{\mbox{\tiny V}}}
\def\DRV{\Delta_{\mbox{\tiny R}}^{\mbox{\tiny V}}}}

\begin{document}


\title{The $\beta$ decay of $^{38}$Ca: Sensitive test of isospin symmetry-breaking corrections from mirror superallowed $0^+$$\rightarrow 0^+$\,transitions} 


\author{H.I. Park}
\email[]{hpark@comp.tamu.edu}

\author{J.C. Hardy}
\email[]{hardy@comp.tamu.edu}

\author{V.E. Iacob}

\author{M. Bencomo}

\author{L. Chen}

\author{V Horvat}

\author{N. Nica}

\author{B.T. Roeder}

\author{E. Simmons}

\author{R.E. Tribble}

\author{I.S. Towner}

\affiliation{Cyclotron Institute, Texas A\&M University, College Station, TX 77845-3366, USA}

\date{\today}

\begin{abstract}
We report the first branching-ratio measurement of the superallowed $0^+$$\rightarrow 0^+$\,$\beta$-transition from $^{38}$Ca.  The result, 0.7728(16), leads
to an  $ft$ value of 3062.3(68)s with a relative precision of $\pm$0.2\%.  This makes possible a high-precision comparison of the $ft$ values for the mirror superallowed
transitions, $^{38}$Ca\,$\rightarrow$$^{38m}$K and $^{38m}$K\,$\rightarrow$$^{38}$Ar, which sensitively tests the isospin symmetry-breaking corrections required to extract
$V_{ud}$, the up-down quark-mixing element of the Cabibbo-Kobayashi-Maskawa (CKM) matrix, from superallowed $\beta$ decay.  The result supports the corrections currently
used, and points the way to even tighter constraints on CKM unitarity.
\end{abstract}

\pacs{23.40.Bw, 27.30.+t, 12.15.Hh}

\maketitle

Superallowed $0^+$$\rightarrow$\,$0^+$\,$\beta$ decay is the experimental source of the most precise value for $V_{ud}$, the up-down quark-mixing element of the
Cabibbo-Kobayashi-Maskawa (CKM) matrix.  According to the standard model, this matrix should be unitary, and currently the most exacting test of that expectation is the
top-row sum, $|V_{ud}|^2$+$|V_{us}|^2$+$|V_{ub}|^2$, which equals 1.00008(56) with the most recent data being used \cite{Ha13}.  Thus, unitarity is satisfied with a small
enough uncertainty ($\pm$0.06\%) to place meaningful constraints on some proposed extensions to the standard model \cite{To10a}.  Any reduction in this uncertainty would
tighten those constraints.

\begin{figure}[b]
\epsfig{file=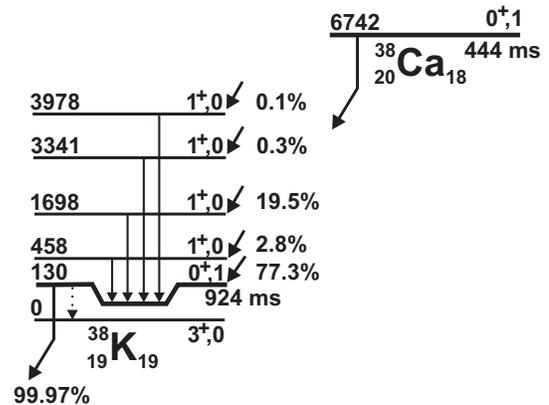, width=7cm}
\caption{Beta-decay scheme of $^{38}$Ca showing the most intense branches.  For each level, its ($J^\pi,T$) is given as well as its energy expressed in keV relative to the
$^{38}$K ground state.  Branching percentages come from this measurement.}
\label{fig1}
\end{figure}

We have measured for the first time precise branching ratios for the $\beta$ decay of $^{38}$Ca (see Fig.\,\ref{fig1}), which includes a superallowed $0^+$$\rightarrow$\,$0^+$
branch not previously characterized.  With the corresponding $Q_{EC}$ value \cite{Er11} and half-life \cite{Bl10,Pa11} already known, the transition's $ft$ value can now be
determined to $\pm$0.2\%.  This is the first addition to the set of well known superallowed transitions \cite{Ha09} in nearly a decade and, being from a $T_Z$\,=\,$-1$ parent
nucleus, it provides the opportunity to make a high-precision comparison of the $ft$ values from a pair of mirror superallowed decays,
$^{38}$Ca\,$\rightarrow$$^{38m}$K and $^{38m}$K\,$\rightarrow$$^{38}$Ar.  The ratio of mirror $ft$ values is very sensitive to the model used to calculate the small
isospin symmetry-breaking corrections that are required to extract $V_{ud}$ from the data.  Since the uncertainty in these corrections contributes significantly
to the uncertainty both on $V_{ud}$ and on the unitarity sum, experimental constraints imposed by mirror $ft$-value ratios can serve to reduce those uncertainties.

Because $0^+$$\rightarrow$\,$0^+$ $\beta$ decay between $T$\,=\,1 analog states depends exclusively on the vector part of the weak interaction, conservation of the vector
current (CVC) requires the experimental $ft$ values to be related to a fundamental constant, the vector coupling constant $G_{\mbox{\tiny V}}$, which is the same for all
such transitions.  In practice, the expression for $ft$ includes several small ($\sim$1\%) correction terms.  It is convenient to combine some of these terms with the
$ft$ value and define a ``corrected" $\F t$ value \cite{Ha09} as follows:
\be
\F t \equiv ft (1 + \delta_R^{\prime}) (1 + \delta_{NS} - \delta_C ) = \frac{K}{2 \GV^2 
(1 + \DRV )}~,
\label{Ftconst}
\ee
where $K/(\hbar c )^6 = 2 \pi^3 \hbar \ln 2 / (m_e c^2)^5 = 8120.2787(11) \times 10^{-10}$ GeV$^{-4}$s and the isospin-symmetry-breaking correction is denoted by
$\delta_C$.  The transition-independent part of the radiative correction is $\Delta_{\mbox{\tiny R}}^{\mbox{\tiny V}}$, while the terms $\delta_R^{\prime}$ and
$\delta_{NS}$ comprise the transition-dependent part of the radiative correction, the former being a function only of the decay energy and the $Z$ of the daughter
nucleus, while the latter, like $\delta_C$, depends in its evaluation on the details of nuclear structure.  From this equation, it can be seen that each measured
transition establishes an individual value for $G_{\mbox{\tiny V}}$ and, according to CVC, all these values -- and all the $\F t$ values themselves -- 
should be identical within uncertainties, regardless of the nuclei involved.  This expectation is strongly supported by the data from all 13 well known
$0^+$$\rightarrow$\,$0^+$ transitions \cite{Ha09}.

Accepting the constancy of $\F t$, we can use Eq.\,(\ref{Ftconst}) to write the ratio of experimental $ft$ values for a pair of mirror superallowed transitions as follows:
\be
\frac{ft^a}{ft^b} = 1+(\delta^{\prime b}_R-\delta^{\prime a}_R)+(\delta^b_{NS}-\delta^a_{NS})-(\delta^b_C-\delta^a_C)~,
\label{ftratio}
\ee
where superscript ``$a$" denotes the decay of the $T_Z$\,=\,$-1$ parent ($^{38}$Ca\,$\rightarrow$$^{38m}$K in the present case) and ``$b$" denotes the decay of the $T_Z$\,=\,0
parent ($^{38m}$K\,$\rightarrow$$^{38}$Ar).  The advantage offered by Eq.\,(\ref{ftratio}) is that the (theoretical) uncertainty on a difference term such as
$(\delta^b_C-\delta^a_C)$ is significantly less than the uncertainties on $\delta^b_C$ and $\delta^a_C$ individually.

To understand this, one must first recognize how $\delta_C$ and its quoted uncertainty were derived in the first place \cite{To08}.  The term itself was broken down into two
components, $\delta_{C1}$ and $\delta_{C2}$, with the first corresponding to a finite-sized shell-model calculation typically restricted to one major shell, while the
second took account of configurations outside that model space via a calculation of the mismatch between the parent and daughter radial wave functions.  The parameters used
for the shell-model calculation were taken from the literature, where they had been based on a wide range of independent spectroscopic data from nearby nuclei.  In all
cases, more than one parameter set was available, so more than one calculated value was obtained for each correction term.  The value adopted for $\delta_{C1}$ was
then the average of the results obtained from the different parameter sets, and the quoted ``statistical" uncertainty reflected the scatter in those results.  If the same
approach is used to derive the mirror differences of correction terms ($\delta^b_{C1}-\delta^a_{C1}$), the scatter among the results from different parameter sets is less
than the scatter in either $\delta^b_{C1}$ or $\delta^a_{C1}$.

For $\delta_{C2}$ there is a further source of theoretical uncertainty that arises from the choice of potential used to obtain the parent and daughter radial wave functions.
Both Woods-Saxon (WS) and Hartree-Fock (HF) eigenfunctions have been used but there is a consistent difference between their results.  Consequently a ``systematic"
uncertainty corresponding to half the difference has been assigned to $\delta_{C2}$, which naturally increases the uncertainty on the derived $V_{ud}$ and on the unitarity
sum \cite{Ha09}.

\begin{table}[t]
\caption{\label{table1}Calculated $ft^a$/$ft^b$ ratios for four doublets with Woods-Saxon (WS) and Hartree-Fock (HF) radial wave functions used in the calculation
of $\delta_C$.}
\vspace{2mm}
\begin{ruledtabular}
\begin{tabular}{lcc}
\multicolumn{1}{l}{Decay pairs, $a$;$b$}&
\multicolumn{2}{c}{$ft^a$/$ft^b$} \\
\cline{2-3}
 & WS & HF \\ 
\hline \\ [-3mm]
$^{26}$Si\,$\rightarrow$$^{26m}$Al ; $^{26m}$Al\,$\rightarrow$$^{26}$Mg & 1.00389(26) & 1.00189(26) \\
$^{34}$Ar\,$\rightarrow$$^{34}$Cl ; $^{34}$Cl\,$\rightarrow$$^{34}$S & 1.00171(26) & 0.99971(43) \\
$^{38}$Ca\,$\rightarrow$$^{38m}$K ; $^{38m}$K\,$\rightarrow$$^{38}$Ar & 1.00196(39) & 0.99976(43) \\
$^{42}$Ti\,$\rightarrow$$^{42}$Sc ; $^{42}$Sc\,$\rightarrow$$^{42}$Ca & 1.00566(65) & 1.00296(42) \\
\end{tabular}
\end{ruledtabular}
\end{table}

With the statistical (theoretical) uncertainty contribution from $\delta_C$ reduced in the mirror $ft$-value ratio, Eq.\,(\ref{ftratio}) offers the opportunity to use
experiment to distinguish cleanly between WS and HF radial wave functions.  If one set of calculations were to be convincingly eliminated, then the systematic uncertainty
on $\delta_C$ could also be eliminated and the uncertainty in $V_{ud}$ reduced.

With current capabilities for producing superallowed $T_Z$\,=\,0 parent nuclei in sufficient quantity for
a high-statistics measurement, there are four mirror pairs that can be completed.  These are listed in Table~\ref{table1}, where it can be seen that the calculated differences between
the WS and HF calculations range from 0.20(2)\% for the mass-26 pair to 0.27(6)\% for mass 42.  Though small, these differences are large enough for experiment to be capable of selecting
one calculation over the other.

We produced 444-ms $^{38}$Ca using a 30$A$-MeV $^{39}$K primary beam from the Texas A\&M K500 superconducting cyclotron to initiate the
$^1$H($^{39}$K, 2$n$)$^{38}$Ca reaction on a LN$_2$-cooled hydrogen gas target.  The fully stripped ejectiles were separated by their
charge-to-mass ratio, $q/m$, in the MARS recoil separator \cite{Tr91}, producing a $^{38}$Ca beam at the focal plane, where the beam composition
was monitored by the periodic insertion of a position-sensitive silicon detector.  With the detector removed, the $^{38}$Ca beam exited the vacuum
system through a 50-$\mu$m-thick Kapton window, passed successively through a 0.3-mm-thick BC-404 scintillator and a stack of aluminum
degraders, finally stopping in the 76-$\mu$m-thick aluminized Mylar tape of a fast tape-transport system.  The combination of $q/m$
selectivity in MARS and range separation in the degraders provided implanted samples that were 99.7\% pure $^{38}$Ca, with the main
surviving trace contaminants being $^{34}$Ar, $^{35}$Ar and $^{36}$K.  Approximately 24,000 atoms/s of $^{38}$Ca were implanted in the tape.

During the measurement, each $^{38}$Ca sample was accumulated in the tape for 1.6 s, with its rate of accumulation being measured by the scintillation
detector located ahead of the degrader stack.  Then the beam was turned off and the tape moved the sample in 200 ms to a shielded counting location 90 cm away,
where data were collected for 1.54~s, after which the cycle was repeated.  This computer-controlled sequence was repeated continuously for nearly
5 days.

\begin{figure}[t]
\epsfig{file=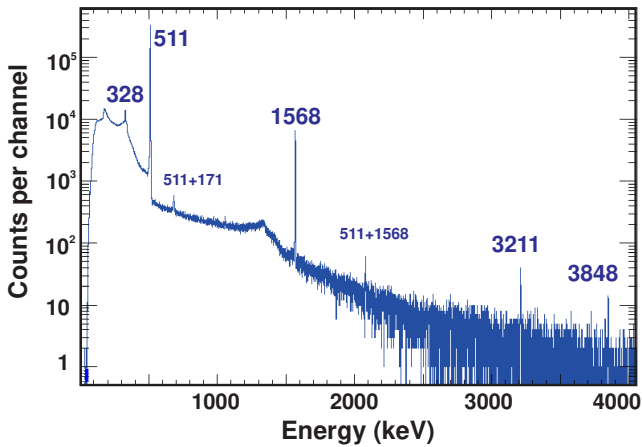, width=8.5cm}
\caption{Spectrum of $\gamma$ rays observed in prompt coincidence with positrons from the decay of $^{38}$Ca.  The small peak labeled ``511+171" is caused by
positron annihilation, from which one 511-keV $\gamma$ ray sums with a back-scattered $\gamma$ ray from the second 511-keV $\gamma$ ray.  The ``511+1568" peak
is the result of coincidence summing between a 1568-keV $\gamma$ ray and annihilation radiation from the positron decay that preceded it.}
\label{fig2}
\end{figure}

At the counting location, the sample was positioned precisely between a 1-mm-thick BC-404 scintillator to detect $\beta^+$ particles, and
our specially calibrated 70\% HPGe detector for $\gamma$ rays.  The former was located 3 mm from one side of the tape, while the latter was 15.1
cm away on the other side.  We saved $\beta$-$\gamma$ coincidences event-by-event, recording the energy of each $\beta$ and $\gamma$ ray, the
time difference between their arrival, and the time that the event itself occurred after the beginning of the counting period.  For each cycle we also
recorded the rate of accumulation of $^{38}$Ca ions in the tape as a function of time, the total number of $\beta$- and $\gamma$-ray singles, and the
output from a laser ranging device that recorded the distance of the stopped tape from the HPGe detector.  From cycle to cycle that distance
could change by a few tenths of a millimeter, enough to require a small adjustment to the HPGe detector efficiency.  Our recorded spectrum of
$\beta$-coincident $\gamma$ rays appears in Fig\,\ref{fig2}.

It can be seen from Fig.\,\ref{fig1} that all $\beta$ transitions from $^{38}$Ca populate prompt-$\gamma$-emitting levels in $^{38}$K, except for the superallowed
branch.  To obtain the superallowed branching ratio, our approach is first to determine the number of 1568-keV $\gamma$ rays relative to the total number of
positrons emitted from $^{38}$Ca.  This establishes the $\beta$-branching ratio to the $1^+$ state in $^{38}$K at 1698 keV.  Next, from the relative
intensities of all the other (weaker) observed $\gamma$-ray peaks, we determine the total Gamow-Teller $\beta$-branching to all $1^+$ states.  Finally, by subtracting
this total from 100\%, we arrive at the branching ratio for the superallowed transition.

More specifically, if the $\gamma$ ray de-exciting state $i$ in $^{38}$K is denoted by $\gamma_i$, then the $\beta$-branching ratio, $R_i$, for the
$\beta$-transition populating that state can be written:
\be
R_i = \frac{N_{\beta \gamma_i}}{N_\beta~\epsilon_{\gamma_i}}~ \frac{\epsilon_\beta}{\epsilon_{\beta_i}} ~,
\label{Ri}
\ee
where $N_{\beta \gamma_i}$ is the total number of $\beta$-$\gamma$ coincidences in the $\gamma_i$ peak; $N_\beta$ is the total number of beta
singles corresponding to $^{38}$Ca $\beta$ decay; $\epsilon_{\gamma_i}$ is the efficiency of the HPGe detector for detecting $\gamma_i$; $\epsilon_{\beta_i}$ is
the efficiency of the plastic scintillator  for detecting the betas that populate state $i$; and $\epsilon_{\beta}$ is the average efficiency for detecting the
betas from all $^{38}$Ca transitions.

{\it Efficiency calibration:} Eq.\,(\ref{Ri}) highlights the importance of having a precise absolute efficiency calibration for the $\gamma$-ray detector, and a
reasonable knowledge of relative efficiencies in the $\beta$ detector.  Our HPGe detector's efficiency has been meticulously calibrated with source measurements
and Monte Carlo calculations to $\pm$0.2\% absolute ($\pm$0.15\% relative) between 50 and 1400 keV \cite{He03}; to $\pm$0.4\% above that, up to 3500 keV \cite{He04};
and to $\pm$1.0\% up to 5000 keV \cite{Me12}.  For the 1568-keV $\gamma$ ray, the peak efficiency is $\epsilon_{\gamma_i}=0.1777(4)$\%.  The relative efficiency of the
plastic scintillator has been determined as a function of $\beta$ energy by Monte Carlo calculations, which have been checked by comparison with measurements on
sources that emit both betas and conversion electrons \cite{Go08}.  For the $\beta$-transition feeding the 1568-keV $\gamma$ ray, the ratio is
$\epsilon_{\beta}$/$\epsilon_{\beta_i}=1.0038(4)$.

{\it Beta singles:} the presence of $N_\beta$ in Eq.\,(\ref{Ri}) makes clear how essential it is to deposit a nearly pure $^{38}$Ca source and to know quantitatively the
weak impurities that remain.  We identified weak contaminant beams at the MARS focal plane, then calculated their energy loss in the degraders, and thus derived the
amount of each that stopped in our tape.  From that, we determined that all impurities contributed only 0.6(3)\% to the total number of betas recorded. 
Much more significant is the contribution from the decay of $^{38m}$K, the daughter of $^{38}$Ca.  This nuclide is not present in the beam, but it naturally
grows in the collected sample as the $^{38}$Ca decays.  Since the half-lives of $^{38}$Ca and $^{38m}$K are well known \cite{Bl10,Pa11,Ha09}, the ratio of their
activities could be accurately calculated, based on the measured time dependence of the $^{38}$Ca deposit rate.  We determined that the betas from $^{38}$Ca constituted
35.10(2)\% of the combined betas from $^{38}$Ca and $^{38m}$K.  Finally, a Monte-Carlo-calculated correction factor of 0.99957(4) was applied to account for $^{38}$Ca
$\gamma$-rays being detected in the thin $\beta$-detector.

{\it Beta-coincident 1568-keV gamma rays:} To obtain the $\beta$-coincident $\gamma$-ray spectrum in Fig.\,\ref{fig2} we gated on the prompt peak in the $\beta$-$\gamma$
time-difference spectrum and subtracted the random-coincidence background.  Our procedure for extracting $\gamma$-peak areas was then to use a modified version of GF3, the
least-squares peak-fitting program in the RADWARE series \cite{Rapc}.  In doing so, we were using the same fitting procedure as was used in the original detector-efficiency
calibration \cite{He03,He04,Me12}.  To determine $N_{\beta \gamma_i}$ for the 1568-keV transition, coincident summing with 511-keV annihilation radiation also had to be
accounted for.  Although the sum peak at 2079 keV could be seen and its area determined, the summing loss from the 1568-keV peak area depends on the total 511-keV response
function: peak plus Compton distribution.  This response function is, in principle, displayed in the first 511 keV of the spectrum in Fig.\,\ref{fig2}; but, in practice, it
is distorted by the response to the 328-keV $\gamma$ ray.  We therefore used an off-line source of $^{22}$Na -- a positron emitter with no $\gamma$ ray below 511 keV -- to
help establish the required response function.  The final summing correction factor for the 1568-keV $\gamma$ ray (including an adjustment for annihilation in flight) is 1.0263(26).  

{\it Correction for dead time and pile-up:}  Dead time in the $\beta$-counting system is small, and affects equally both the numerator and denominator in Eq.\,(\ref{Ri}), so it
does not influence our result.  However, dead time and pile up do affect the much slower signals from the HPGe detector, and they depend not only on the rate of
coincident $\gamma$ rays, which averaged 94 counts/s, but also on the singles $\gamma$ rate, which averaged 430 counts/s.  Furthermore, the rate during each cycle also
decreased with time.  Taking account of these effects we arrived at a dead-time/pile-up correction factor of 1.01366(11).

Combining these results into Eq.\,(\ref{Ri}), we found the branching ratio for the $\beta$ transition to the 1698-keV state in $^{38}$K to be 0.1949(13).  Then, by analyzing the
full $\gamma$-ray spectrum of Fig.\,\ref{fig2}, and making provision for weak $1^+$$\rightarrow$$1^+$ $\gamma$ transitions, we obtained the total of all Gamow-Teller branches
relative to this transition.  In this process, account had to be taken of the
small electron-capture competition with $\beta^+$ decay, since the former would not have led to a coincidence in our spectrum. This only has a slight impact on the two lowest-energy
(and weakest) $\beta$ transitions.  Our final result for the total Gamow-Teller branching from $^{38}$Ca is 0.2272(16), and this leads to a superallowed branching ratio of
$0.7728 \pm 0.0014_{stat} \pm 0.0009_{syst}$ or, with the uncertainties combined in quadrature, $0.7728 \pm 0.0016$.  The full details of this experiment and its analysis will appear
in a subsequent publication \cite{Pa14}.

\begin{figure}[t]
\epsfig{file=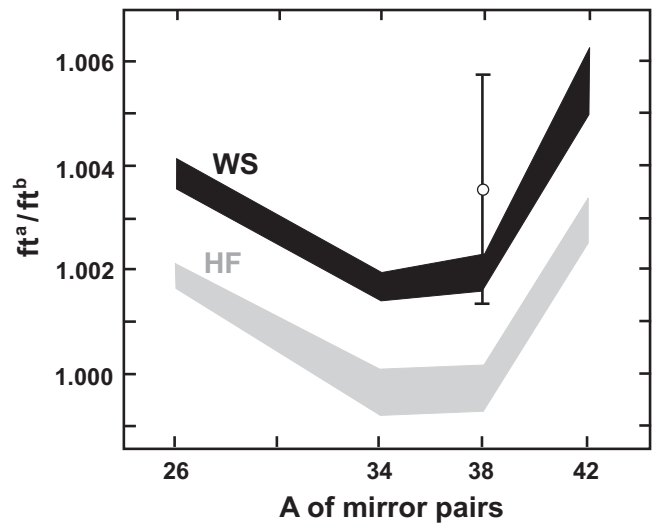, width=8.5cm}
\caption{Mirror-pair $ft^a/ft^b$ values for $A$ = 26, 34, 38 and 42, the four cases currently accessible to high-precision experiment. The black and grey bands connect calculated
results that utilize Woods-Saxon (WS) and Hartree-Fock (HF) radial wave functions, respectively (see Table\,\ref{table1}).  Our measured result for the $A=38$ mirror pair is shown as
the open circle with error bars. }
\label{fig3}
\end{figure}

The half-life of $^{38}$Ca is 443.77(35)\,ms \cite{Bl10,Pa11} and the $Q_{EC}$ value for its superallowed branch is 6612.12(7)\,keV \cite{Er11}.  Taking these results with
our new value for the branching ratio and correcting for electron capture, we arrive at an $ft$ value for the $^{38}$Ca superallowed branch of $ft^a = 3062.3(68)$\,s.  The $ft$ value for the
mirror transition from $^{38m}$K is $ft^b = 3051.5(9)$\,s, a value that comes from the 2009 survey \cite{Ha09} updated for a more recent $Q_{EC}$ measurement reported by Eronen {\it et al.}
\cite{Er09}.  The ratio of the two, $ft^a/ft^b = 1.0036(22)$, appears in Fig.\,\ref{fig3}, where it can be compared with the calculated results from Table\,\ref{table1}.

Although our experimental result favors the WS calculation, it is not yet definitive.  Nevertheless, it clearly points the way to a potential reduction of the uncertainty on $V_{ud}$
through the elimination of alternatives to the WS-calculated corrections currently used.  The lack of precise branching-ratio measurements has so far prevented the $T_Z=-1$ decays of $^{26}$Si, $^{34}$Ar,
$^{38}$Ca and $^{42}$Ti from being fully characterized with high precision.  Now that we have demonstrated the capability to make such a measurement on $^{38}$Ca, the other three
cases should not be far behind.  Together, if all four convey a consistent message, they can have a major impact by sensitively discriminating among the models used to calculate the
isospin-symmetry-breaking corrections.

The precision that could be quoted here for the superallowed transition benefited from the fact that the branching ratios actually measured were significantly smaller than the superallowed
one, which was derived by subtraction of the measured values from 1.  This had a very salutary effect on the relative uncertainty for the superallowed branch, reducing it by a multiplicative factor of
0.3 (= 0.227/0.773) compared to the measured Gamow-Teller branches.  As to the other three mirror cases: $^{42}$Ti has no such reduction effect, but for $^{26}$Si the reduction factor is
again 0.3, and for $^{34}$Ar it is an impressive 0.06.  Currently, there is a 40-year-old measurement \cite{Ha74} of the total Gamow-Teller branches from $^{34}$Ar, which has a relative
uncertainty of 4.5\%, far higher than we have demonstrated possible today.  Even with the reduction factor, the $ft^a/ft^b$ ratio it provides for $A=34$ has a larger uncertainty than
our $A=38$ result and is not useful for the Fig.\,\ref{fig3} comparison with theory.  However, with a new branching-ratio measurement employing the techniques described here, a very tight
uncertainty should be anticipated for the $A=34$ $ft$-value ratio.  We are currently well advanced in making such a measurement.  It may also be noted that the result presented here was limited
to a relative uncertainty of 0.21\% by the counting statistics acquired during less than 5 days of accelerator beam time.  Ultimately, with additional running time the overall relative
uncertainty for this $A=38$ case could be reduced towards the limit of 0.12\% set by systematic effects.

As a final remark, we point out that results such as these, which are at the limits of experimental precision, benefit enormously from repetition by independent groups.  The robustness of
the current data set for $0^+$$\rightarrow 0^+$\,superallowed $\beta$ decay can be attributed to the multiple measurements that contribute to each input datum.  These branching ratios
for the $T_Z=-1$ parent nuclei should not stand as lingering exceptions.

\begin{acknowledgments}

This work was supported by the U.S. Department of Energy under Grant No.\,DE-FG03-93ER40773 and by the Robert
A. Welch Foundation under Grant No.\,A-1397.

\end{acknowledgments}

\end{document}